\documentclass[]{aa}

\usepackage{psfig}

\newcommand\pp     {$\pm$}
\newcommand\pers     {s$^{-1}$}
\newcommand\micros  {$\mu$s}

\begin{document}

\thesaurus{06 (02.01.2;  08.09.2;
               08.14.1; 13.25.5)}

\title{The X-ray Timing Behavior of the X-ray Burst Source SLX
1735--269}

\author{Rudy Wijnands \& Michiel van der Klis}

\authorrunning{Wijnands \& van der Klis}
\titlerunning{X-ray Timing Behavior of SLX 1735--269}

\offprints {Rudy Wijnands, email: rudy@astro.uva.nl}

\institute{Astronomical Institute ``Anton Pannekoek'' and Center for High
Energy Astrophysics, University of Amsterdam, Kruislaan 403, NL-1098
SJ Amsterdam, The Netherlands; rudy@astro.uva.nl,
michiel@astro.uva.nl}

\date{Received; accepted}

\maketitle 



\begin{abstract}

We report for the first time on the rapid X-ray variability of the
galactic bulge source and X-ray burster SLX 1735--269. The power
spectrum as observed with the {\it Rossi X-ray Timing Explorer} is
characterized by a strong band-limited noise component which is
approximately flat below a 0.1--2.3 Hz break frequency; above this
frequency the power spectrum declines as a power law of index 0.9. At
the highest observed count rate a broad bump is superimposed on this
band-limited noise.  The power spectrum is very similar to that of
other low-luminosity neutron-star low-mass X-ray binaries (LMXBs) and
to black-hole candidates when these types of source accrete at their
lowest observed mass accretion rates. However, we identify one unusual
aspect of the X-ray variability of SLX 1735--269: the break frequency
increases when the inferred mass accretion rate decreases. This is the
opposite to what is normally observed in other sources. The only
source for which the same behavior has been observed is the
accretion-powered millisecond X-ray pulsar SAX J1808.4--3658. No
coherent millisecond pulsations were observed from SLX 1735--269 with
an upper limit on the amplitude of 2.2\% rms.  Observing this behavior
in SLX 1735--269 increases the similarities between SAX J1808.4--3658
and the other neutron star LMXBs for which so far no coherent
pulsations have been observed. We expect that other sources will show
the same behavior when these sources are studied in detail at their
lowest mass accretion rates.


\end{abstract}

\newcommand{\bdouble}{\baselineskip 1.5\baselineskip}
\newcommand{\edouble}{\par \baselineskip .66667\baselineskip}

\section{Introduction \label{intro}}

The galactic bulge source SLX 1735--269 was discovered in 1985 by
Skinner et al. (1987) during the {\it Spacelab 2} mission. Although
observed on several occasions with other X-ray instruments (e.g., {\it
GRANAT}/SIGMA: Goldwurm et al. 1996 and references therein; {\it
ASCA}: David et al. 1997), little is known about this source. Goldwurm
et al. 1996 detected the source up to about 150 keV with a spectral
index above 30 keV of $\sim-3$. This is steeper than usually observed
for black-hole candidates and therefore they tentatively suggested
that the compact object in the system is a neutron star. {\it ASCA}
observations of this source below 10 keV also could not uniquely
identify the nature of the compact object (David et al. 1997) but they
were consistent with the neutron star hypothesis.  The issue of the
nature of the compact object in SLX 1735--269 was finally settled by
the discovery of a type I X-ray burst from this source using the Wide
Field Cameras onboard {\it BeppoSAX} (Bazzano et al. 1997a; Bazzano et
al. 1997b; Cocchi et al. 1998), demonstrating that SLX 1735--269 is a
low-mass X-ray binary (LMXB) containing a neutron star. 

So far, the rapid X-ray variability of this source has not been
studied in detail.  The neutron star nature of this system motivated
us to analyze the timing behavior of this source as observed by the
{\it Rossi X-ray Timing Explorer} ({\it RXTE}). We searched for
quasi-periodic oscillations (QPOs) between 300 and 1200 Hz, which are
often observed in neutron star LMXBs (see van der Klis 1998, 1999 for
reviews), and coherent pulsations such as observed in the
accretion-driven millisecond X-ray pulsar SAX J1808.4--3658 (Wijnands
\& van der Klis 1998a). Although those phenomena were not detected, we
discovered one characteristic of the timing behavior which, so far,
has only been observed for SAX J1808.4--3658, increasing the
similarity between SAX J1808.4--3658 and the other neutron star LMXBs.

\section{Observations, analysis, and results \label{results}}

SLX 1735--269 was observed using {\it RXTE} on several occasions (see
Table~\ref{obslog} for a log of the observations) for a total of 11
ksec of on-source data. Data were collected simultaneously with 16 s
time resolution in 129 photon energy channels (effective energy range
2--60 keV), and with 1 \micros~time resolution in 256 channels (2--60
keV). A light curve, an X-ray color-color diagram (CD), and an X-ray
hardness-intensity diagram (HID) were created using the 16 s data, and
power spectra (for the energy range 2--60 keV) were calculated from
the 1~\micros~data using 256 s FFTs.

\begin{figure}[t]
\begin{center}
\begin{tabular}{c}
\psfig{figure=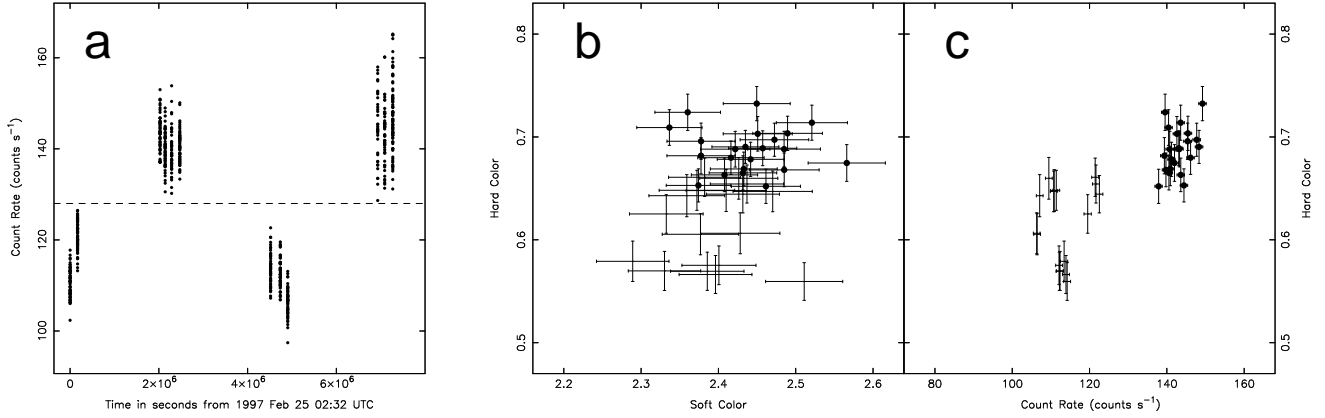,width=18cm}
\end{tabular}
\caption{The 2.0--16.0 keV light curve ({\it a}), the color-color
diagram (CD) ({\it b}), and the hardness-intensity diagram (HID) ({\it
c}) of SLX 1735--269. All count rates were background
subtracted. Dead-time corrections are negligible.  The dashed line at
128 counts \pers~in {\it a} indicates the two different count rate
ranges used to select the power spectra. In the CD, the soft color is
the count rate ratio between the 3.5--6.4 and 2.0--3.5 keV bands; the
hard color that between 9.7--16.0 and 6.4--9.7 keV. In the HID, the
hard color is the same as in the CD and the count rate is the count
rate in the photon energy range 2.0--16.0 keV. In {\it a} the points
are 16 s averages; in {\it b} and {\it c} the points are 256 s
averages. The errors on the count rate in {\it a} are typically
2\%--4\%. In {\it b} and {\it c}, the filled dots are the data above a
count rate of 128 counts \pers~and the other points the data below
this value.
\label{lightcurve}}
\end{center}
\end{figure}

Figure~\ref{lightcurve} shows the background subtracted (using
PCABACKEST version 2.1b and the faint sources L7/240 background model)
light curve (2.0--16.0 keV; all 5 detectors on;
Fig.~\ref{lightcurve}{\it a}), the CD (Fig.~\ref{lightcurve}{\it b}),
and the HID (Fig.~\ref{lightcurve}{\it c}) (for the energy bands used
to calculate the colors used in those diagrams see the caption of the
figure). The 2.0--16.0 keV count rate varies between $\sim$100 and
$\sim$160 counts \pers~(Fig.~\ref{lightcurve}{\it a} and {\it c}). The
hard color tends to increase when the count rate increases
(Fig.~\ref{lightcurve}{\it c}); the soft color does not have a clear
correlation with count rate.

\begin{figure}[t]
\begin{center}
\begin{tabular}{c}
\psfig{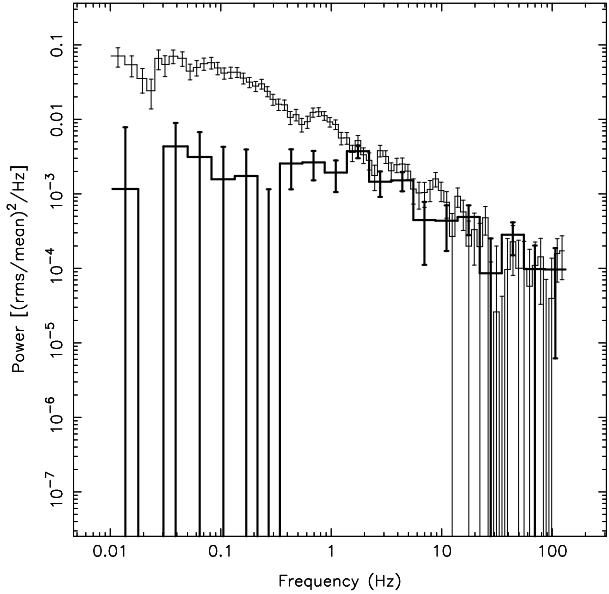}
\end{tabular}
\caption{The power spectra (2--60 keV) obtained for SLX 1735--269. The
thin line is the power spectrum corresponding to a count rate of $>$
128 counts \pers~(2.0--16.0 keV; background subtracted) and the thick
line is the power spectrum corresponding to a count rates of $<$ 128
counts \pers~(see also Fig.~1{\it a}). The spectra were
logarithmically rebinned and the Poisson level was subtracted.  Due to
lower count rates, less data, and weaker band-limited noise (see
Table~\ref{fitparameters}), the signal-to-noise ratio of the power
spectrum corresponding to the low count rate selection was much less
than that of the other power spectrum. For display purposes, different
frequency bin sizes were used for the two power spectra.
\label{powerspectra}}
\end{center}
\end{figure}

We selected power spectra based on the background corrected count
rates in the 2.0--16.0 keV band and averaged them. We obtained two
average power spectra: the first one corresponds to a count rate
$<$128 counts \pers; the second one to $>$128 counts \pers~(see also
Fig~\ref{lightcurve}{\it a}).  These two power spectra are shown in
Figure~\ref{powerspectra}. Although the count rate differs only
slightly between the two selections (10\%--20\%) the difference in the
power spectra is remarkable.  Both power spectra show a broad
band-limited noise component. They were fitted (after subtraction of
the Poisson level) with a broken power law. For the high count rate
power spectrum also a Lorentzian was added in order to fit the bump
superimposed on the band-limited noise near 0.9 Hz.

The fit parameters are presented in Table~\ref{fitparameters}.  The
break frequency was higher (2.3 Hz) when the count rate was low than
when it was higher, probably by an order of magnitude (see also
Fig.~\ref{powerspectra}). The index below the break during the highest
count rates is $\sim0$; during the lowest count rates this parameter
had to be fixed to 0 due to the low statistics. The indices above the
break in both count rate regimes are consistent with each other at
$\sim$0.9. The strength of the noise is $\sim$24\% rms in the high
count rate selection, and $\sim$17\% in the low count rate selection.
The bump present on top of the broad-band noise at high count rate had
an amplitude of 4.7 \% rms (3.3$\sigma$), a FWHM of 0.3~Hz, and a
frequency of 0.87~Hz.

We searched for kHz QPOs but none were found, with upper limits (95 \%
confidence levels) between 13 to 26 \% rms (depending on count rate
selection, frequency range, and assumed FWHM of the kHz QPO). These
upper limits are higher than the strengths of kHz QPOs detected in
other low-luminosity neutron star LMXBs. Therefore, we cannot exclude
the presence of QPOs with frequencies between 100 and 1500 Hz. Upper
limits (95 \% confidence level) on coherent pulsations in the
frequency range 100--1000 Hz were 2.2\% rms, which is significantly
lower than the 4\%--6 \% rms detected for the accretion-driven
millisecond X-ray pulsar SAX J1808.4--3658 (Wijnands \& van der Klis
1998a; Cui, Morgan, \& Titarchuk 1998). However, it is possible that
SAX J1808.4--3658 has a low system inclination (see Chakrabarty \&
Morgan 1998) and that SLX 1735--269 has a much larger inclination. The
pulsations in SLX 1735--269 will then be smeared out over many
frequency bins, making a 4\%--6\% rms amplitude pulsation undetectable
in our analysis.

We fitted the X-ray spectra corresponding to the two power spectral
selections. The X-ray spectrum corresponding to a count rate of $>$128
counts \pers~could be adequately fitted with an absorbed power law
with a power law index of $\sim$2.2 (using an N$_{\rm H}$ of
$1.47\times10^{22}$ atoms cm$^{-2}$; David et al. 1997). The 3--25 keV
flux was $3.8\times10^{-10}$ ergs cm$^{-2}$ \pers, corresponding to an
intrinsic luminosity of $3.3\times10^{36}$ ergs \pers~(assuming a
distance of 8.5 kpc). The X-ray spectrum corresponding to a count rate
of $<$128 counts \pers~was fitted with an absorbed power law with
index 2.4. The fit was considerably improved when a gaussian line,
near 6.7 keV with a width of 0.8 keV, was added. The 3--25 keV flux
was $2.8\times10^{-10}$ ergs cm$^{-2}$ \pers, corresponding to an
intrinsic luminosity of $2.4\times10^{36}$ ergs \pers. These fluxes
are in the range previously observed for SLX 1735--269. The steeper
power law for the low count rate selection is consistent with the
smaller hard color in the CD (Fig.~\ref{lightcurve} {\it left}),
compared to that of the high count rate selection.

\section{Discussion \label{discussion}}

We presented for the first time an analysis of the rapid X-ray
variability properties of the X-ray burst source SLX 1735-269. The
timing properties are very similar to those of other low luminosity
low magnetic field strength neutron star LMXBs and black hole
candidates during their lowest observed mass accretion rates (Wijnands
\& van der Klis 1999 and references therein). The power spectrum is
dominated by a broad band-limited noise component which follows
roughly a power law at high frequency, but breaks at a certain
frequency below which the power spectrum is approximately flat.  When
the statistics are sufficient, a broad bump can be detected super
imposed on top of this band-limited noise, which is also often
observed in other X-ray binaries (see Wijnands \& van der Klis 1999
and references therein). The power spectra resemble those obtained for
the low-luminosity neutron star LMXBs called the atoll sources, when
they accrete at their lowest observed mass accretion rates (i.e., when
they are in their so-called island state). We therefore suggest that
SLX 1735--269 was during the {\it RXTE} observations in the island
state, assuming it is an atoll source.  Also similar to other X-ray
binaries is that when the break frequency changes the high frequency
part above the break remains approximately the same (see, e.g.,
Belloni \& Hasinger 1990).

However, we observe one uncommon feature of the broad-band noise
component: the break frequency increased when the X-ray flux, and
therefore possibly the mass accretion rate, decreased. In atoll
sources the break frequency decreases when the inferred mass accretion
rate decreases (e.g., Prins \& van der Klis 1998; M\'endez et
al. 1997; Ford \& van der Klis 1998). So far, only one other source is
known for which the break frequency has been observed to increase with
decreasing inferred mass accretion rate: the accretion-driven
millisecond X-ray pulsar SAX J1808.4--3658 (see Wijnands \& van der
Klis 1998b). During the beginning of the decay of the 1998 April
outburst of this transient source, the break frequency decreased with
decreasing X-ray flux. However, half-way the decay the break frequency
suddenly increased again while the X-ray flux kept on
decreasing. Wijnands \& van der Klis (1998b) tentatively proposed that
this unexpected behavior of the break frequency could be due to the
unique pulsating nature of SAX J1808.4--3658 compared to the
non-pulsating neutron star LMXBs and black holes candidates, or it
could be due to the first ever detailed study of the timing properties
of a neutron star LMXB at such low mass accretion rates. With our
analysis of SLX 1735--269, which is a persistent LMXB and for which no
coherent millisecond pulsations could be detected, it has been shown
that the latter is most likely the case.  Thus, the unexpected
behavior of the accretion-driven millisecond X-ray pulsar is not a
unique feature of this system, increasing the similarities of that
source with the other, non-pulsating neutron star LMXBs.

\begin{figure}[t]
\begin{center}
\begin{tabular}{c}
\psfig{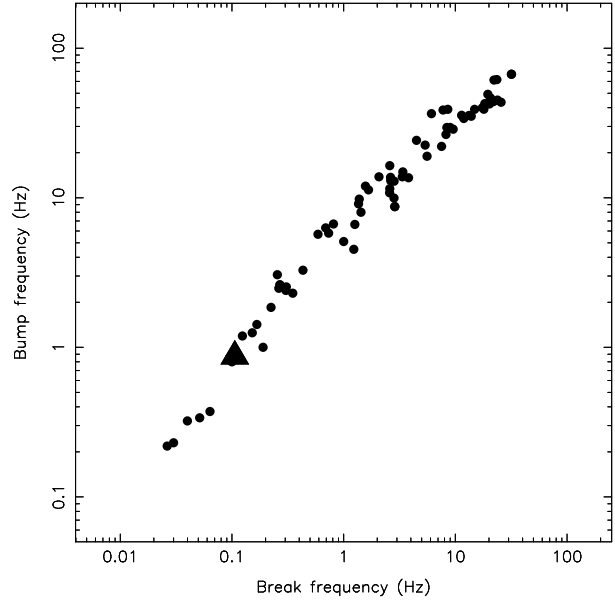}
\end{tabular}
\caption{The frequency of the bump versus the break frequency of the
band-limited noise for several sources. The filled bullets are taken
from Figure 2{\it a} of Wijnands \& van der Klis (1999) and represent
the neutron star LMXBs (including the accretion-driven millisecond
X-ray pulsar SAX J1808.4--3658) and the black-hole candidates analyzed
by them. The filled large triangle is SLX 1735--269. The error bars
are smaller than the size of the data points.
\label{break_vs_bump}}
\end{center}
\end{figure}

During the highest count rates a bump is present on top of the
band-limited noise. Wijnands \& van der Klis (1999) showed that the
frequency of this bump correlates well to the frequency of the break
in low-luminosity neutron star LMXBs (including the accretion-driven
millisecond X-ray pulsar) and black hole
candidates. Figure~\ref{break_vs_bump} shows the same data plotted in
Figure 2{\it a} of Wijnands \& van der Klis (1999), but now including
the data point of SLX 1735--269 (triangle). SLX 1735--269 is right on
the relation defined by the other sources. Again, SLX 1735--269 is
very similar to other low-luminosity LMXBs. The point obtained for SLX
1735--269 is at the low end of the neutron star points (the
lower-frequency points are mostly for black-hole candidates) and very
similar to the data of the X-ray bursters 4U 1812--12 and 1E
1724--3045 (see Wijnands \& van der Klis 1999). The latter two sources
are, therefore, good candidates to display the same increase of the
break frequency with decreasing mass accretion rate.

It is also interesting to note that the 3--25 keV luminosity of SLX
1735--269 ($\sim$2--3 $\times10^{36}$ ergs \pers) is very close to the
3--25 keV luminosity of SAX J1808.4--3658 when in this source the
break-frequency and the mass accretion rate became anti-correlated
($\sim$2$\times10^{36}$ ergs \pers; Wijnands \& van der Klis 1999). It
is possible that this anti-correlation occurs at a specific X-ray
luminosity, which might be similar in all neutron star LMXBs. This can
easily be checked by studying neutron star LMXBs in detail at such low
luminosities.

\begin{acknowledgements}
This work was supported in part by the Netherlands Foundation for
Research in Astronomy (ASTRON) grant 781-76-017 and by the Netherlands
Researchschool for Astronomy (NOVA).  This research has made use of
data obtained through the High Energy Astrophysics Science Archive
Research Center Online Service, provided by the NASA/Goddard Space
Flight Center.

\end{acknowledgements}

\clearpage

\begin{table*}[t]
\caption[]{Log of the observations \label{obslog}}
\begin{flushleft}
\begin{tabular}{ccc}
\hline
Obs. ID & Date      & Start -- End   \\
        & (1997) & (UTC) \\
\hline
20170-03-01-00 & 25 Feb &  02:32--02:50   \\
20170-03-02-00 & 27 Feb &  00:58--01:16   \\
20170-03-03-00 & 20 Mar &  13:16--13:42   \\
20170-03-04-00 & 21 Mar &  21:27--21:43   \\
20170-03-05-00 & 23 Mar &  13:50--14:15   \\
20170-03-06-00 & 25 Mar &  16:57--17:13   \\
20170-03-07-00 & 18 Apr &  09:19--09:40   \\
20170-03-08-00 & 20 Apr &  22:20--22:36   \\
20170-03-09-00 & 22 Apr &  20:46--21:01   \\
20170-03-10-00 & 16 May &  07:58--08:28   \\
20170-03-11-00 & 18 May &  03:38--04:29   \\
20170-03-12-00 & 20 May &  06:06--06:48   \\
\hline
\end{tabular}
\end{flushleft}
\end{table*}

\begin{table*}[t]
\caption[]{Power spectral fit parameters \label{fitparameters}}
\begin{flushleft}
\begin{tabular}{llllllllllll}
\hline
                   &       & \multicolumn{4}{c}{Band limited noise} &&\multicolumn{3}{c}{Bump} & &  \\ \cline{3-6}\cline{8-10}
Count rate$^a$     & \#$^b$ & Break frequency & Rms$^c$ & $\alpha_{\rm below}^d$ & $\alpha_{\rm above}^d$ && Rms$^e$ & FWHM & Frequency & $\chi^2$& d.o.f. \\
(counts \pers)     &       & (Hz)           & (\%)&            &            && (\%) & (Hz) &(Hz)& &    \\
\hline
113\pp6 & 15 & 2.3$^{+1.0}_{-0.7}$ & 16.7\pp1.7 & 0$^f$ & 0.9\pp0.2 & &$<$7$^g$ &  &  &45 & 36 \\
141\pp5 & 21 & 0.11\pp0.02 & 23.6\pp0.6 & $-0.04^{+0.13}_{-0.15}$ &0.91\pp0.03&& 4.7$^{+0.9}_{-0.7}$ & 0.3\pp0.1 & 0.87\pp0.05 & 151 & 162\\
\hline
\multicolumn{12}{l}{$^a$ 2.0--16.0 keV; errors are the standard
deviation of each selection}\\
\multicolumn{12}{l}{$^b$ Number of power spectra averaged}\\
\multicolumn{12}{l}{$^c$ 2--60 keV; integrated over 0.01--100 Hz}\\
\multicolumn{12}{l}{$^d$ $\alpha_{\rm below}$ and $\alpha_{\rm above}$ are the power law
index below and above the break frequency, respectively}\\
\multicolumn{12}{l}{$^e$ 2--60 kev}\\
\multicolumn{12}{l}{$^f$ Parameter fixed}\\
\multicolumn{12}{l}{$^g$ Assuming a FWHM between 5 and 10 Hz and a frequency between 10 and 15 Hz}\\
\end{tabular}
\end{flushleft}
\end{table*}
\clearpage

\clearpage

\end{document}